\documentclass[conference,letterpaper, 10pt]{IEEEtran}

\usepackage[utf8]{inputenc} 
\usepackage[T1]{fontenc}
\usepackage{url}
\usepackage{ifthen}
\usepackage{cite}
\usepackage[cmex10]{amsmath} 

\usepackage[english]{babel}
\usepackage{amssymb}

\usepackage{enumitem}
\usepackage{empheq}
\usepackage{graphicx}
\usepackage[colorinlistoftodos]{todonotes}

\usepackage[short]{optidef}

\RenewEnviron{BaseMiniExclam}[7]{%
  \selectConstraintMult{#1}%
  \renewcommand{\localOptimalVariable}{#2}%
  \begin{subequations}
  \ifthenelse{\equal{#7}{b}}{\allowdisplaybreaks}%
  #4
  \begin{alignat}{5}
    \bodyobj{#2}{#3}{#6}{#5}
    \BODY
  \end{alignat}
  \end{subequations}%
  \setStandardMini
}

\usepackage{mathtools}  
\mathtoolsset{showonlyrefs}  
\allowdisplaybreaks

\def\R{\mathbb{R}}

\def\e{\varepsilon}

\def\bb{\boldsymbol\beta}
\def\bt{\boldsymbol\theta}

\def\obt{\overline{\boldsymbol\theta}}
\def\ot{\overline{\theta}}

\def\e{\varepsilon}

\newtheorem{lem}{Lemma}
\newtheorem{thm}{Theorem}

\newtheorem{cor}{Corollary}
\newtheorem{fact}{Fact}

\def\Rb{R^{\text{bal}}}

\title{Linear Programming Bounds for\\ Almost-Balanced Binary Codes}

\author{%
  \IEEEauthorblockN{Venkatesan Guruswami and Andrii Riazanov}
  \IEEEauthorblockA{Carnegie Mellon University\\
                    Computer Science Department\\
                    Pittsburgh, PA 15213\\
                    Email: \{venkatg,\ riazanov\}@cs.cmu.edu}
}

\begin{document}

\maketitle

\begin{abstract}
We revisit the linear programming bounds for the size vs. distance trade-off for binary codes, focusing on the bounds for the almost-balanced case, when all pairwise distances are between $d$ and $n-d$, where $d$ is the code distance and $n$ is the block length.
We give an \emph{optimal} solution to Delsarte's LP for the almost-balanced case with large distance $d \geq (n - \sqrt{n})/2 + 1$, which shows that the optimal value of the LP coincides with the Grey-Rankin bound for self-complementary codes. 

We also show that a limitation of the asymptotic LP bound shown by Samorodnitsky, namely that it is at least the average of the first MRRW upper bound and Gilbert-Varshamov bound, continues to hold for the almost-balanced case.



\end{abstract}

\section{Introduction}

This paper concerns the standard size-versus-distance tradeoff for binary error-correcting codes. A binary code $C$ with block length $n$ and distance $d$ is a subset of the $n$-dimensional Hamming cube $\{0, 1\}^n$, such that the Hamming distance between any two elements from $C$ is at least $d$. In this paper such codes will be referred to as min-distance codes, and we denote $A(n, d)$ to be the maximal size of such a code. Understanding the behavior of $A(n,d)$ is one of the most fundamental challenges in combinatorial coding theory. For the case when $d$ is linear in $n$, one is interested in asymptotic rates of the code rather than their sizes:
 \vspace{-4pt}
\begin{align}
 d,
\end{align}
where $\delta \approx d/n \leq 1/2$ is the relative distance of the code. 

The best known asymptotic bounds as well as finite-length bounds on $A(n,d)$ for many parameters are given by Delsarte's \emph{linear programming approach}~\cite{Delsarte}. 
Specifically, McEliece, Rodemich, Rumsey, and Welch~\cite{MRRW} obtain the  best known upper bounds on $R(\delta)$ for the entire region $\delta\in (0, 1/2)$ by finding a good feasible solution to Delsarte's LP. There is, however, a sizeable gap between the MRRW upper bound on $R(\delta)$ and the Gilbert-Varshamov lower bound $R(\delta) \ge 1 - h(\delta)$ where $h(\cdot)$ is the binary entropy function, and shrinking this gap has remained a major challenge for almost 45 years. 

In this paper, we are interested in the performance of the linear programming approach for bounding the size of 
\emph{almost-balanced} codes, which obey the stronger condition that the Hamming distance between any two distinct codewords lies between $d$ and $(n-d)$.  Almost-balanced codes when $d=(1-\epsilon)n/2$ are closely related to $\epsilon$-biased spaces that are of fundamental interest in pseudorandomness and derandomization, starting with the seminal work of Naor and Naor~\cite{NN93}. The recent breakthrough explicit construction of high distance codes approaching the Gilbert-Varshamov bound by Ta-Shma crucially proceeds by constructing almost-balanced codes~\cite{TaShma}. 

Almost-balanced codes are closely related to \emph{self-complementary} codes. The code $C'$ is self-complementary if $u \in C'$ implies that the complementary vector $\overline{u}$ is also in $C'$. The connection between almost-balanced and self-complementary codes is two-way:
\begin{itemize}
    \item any  almost-balanced code $C$ with distance $d$ (i.e. all pairwise distances lie in the interval $[d, n-d]$) corresponds to a self-complementary code of size $2|C|$. Indeed, the code $C' = \{ u\,|\, u\text{ or } \overline{u} \text{ is in } C\}$ is self-complementary and has size $2|C|$.
    \item any self-complementary code $C'$ with distance $d$ corresponds to a family of almost-balanced codes of size $|C'|/2$ and distance $d$. Such codes can be constructed by taking one codeword from every complementary pair $u, \overline{u} \in C'$.
\end{itemize}
Thus to translate the bounds for almost-balanced codes to self-complementary codes, one just needs to multiply the size-versus-distance trade-off (for the same distance $d$) by $2$.

Let us denote by $B(n,d)$ the maximal size of an almost-balanced code with block length $n$ and distance $d$. 
Once again, when $d$ is linear in $n$, we can study the asymptotic rate of the codes rather than their sizes:
 \vspace{-4pt}
\begin{align}
 \Rb(\delta) &= \limsup\limits_{n\to\infty} \frac{\log B(n, \lfloor\delta n\rfloor)}{n} \ .
 \end{align}

The driving force behind this work is to study upper bounds obtained for almost-balanced codes via the linear programming method, and compare them to the ones for the min-distance codes. Our results consists of two parts:

\begin{enumerate}
    \item For almost-balanced codes with large distances ${d \geq \frac{n - \sqrt{n}}{2} + 1}$ we find an \emph{optimal solution} to Delsarte's linear program. This solution gives an upper bound on $B(n,d)$ equivalent to the Gray-Rankin bound for self-complementary codes~\cite{grey, rankin, rankin2, McWSloane}. Since the solution we obtain is optimal, this proves a matching \emph{lower bound} on Delsarte's LP problem, showing that LP approach cannot prove better bounds for the almost-balanced case. While this was already implied for a very limited range of pairs $(n, d)$ for which Grey-Rankin bound is achievable (\hspace{1sp}\cite{McGuire}), our bound works for arbitrary even $n, d$ satisfying ${d \geq \frac{n - \sqrt{n}}{2} + 1}$.
    
    \item We show (Theorem~\ref{thm:bal_Samor}) that the asymptotic LP bound for almost-balanced codes is at least the average of the first MRRW bound and Gilbert-Varshamov lower bound. This is the analog of Samorodnitsky's result~\cite{Samorodnistky} for min-distance codes, and indicates that the (direct) linear programming approach cannot attain the best known lower bound on $\Rb(\delta)$. The proof is a simple adaptation of Samorodnitsky's argument.
\end{enumerate}

\section{Krawtchouk polynomials and Delsarte's LP}
For fixed $n$, the Krawtchouk polynomials are defined as 
\begin{equation}
    \label{eq:def_K}
    K_s(x) = \sum_{j=0}^s(-1)^j\binom{x}{j}\binom{n-x}{s-j}, \qquad s = 0, 1, \dots, n.
\end{equation}
$K_s(x)$ is a degree-$s$ polynomial, and $K_0$, $K_1$, $\dots$, $K_n$ form an orthogonal family with respect to a measure $\mu_i = \binom{n}{i}/2^n$:
\begin{equation}
    \label{eq:orthog}
    \sum_{j=0}^n \frac{\binom{n}{j}}{2^n} K_s(j) K_t(j) = \delta_{st}\binom{n}{s}.
\end{equation}
Notice that $K_0(x) \equiv 1$ and $K_i(0) = \binom{n}{i}$. Below are some of the properties of Krawtchouk polynomials we will need.

\begin{fact}
\label{cor:parity}
$K_s(n/2 - x)$ is an even function for even $s$, and is an odd function for odd $s$.
\end{fact}
\begin{fact}
\label{cor:symmetry}
$K_{n-s}(i) = K_s(i)$ for even $i$, and ${K_{n-s}(i) = -K_s(i)}$ for odd $i$.
\end{fact}
\begin{fact}
\label{cor:reciprocity}
Reciprocity property: $\binom{n}{s}K_i(s) = \binom{n}{i}K_s(i)$.
\end{fact}

The Delsarte's upper bound on $A(n, d)$ can be formulated as the following Linear Programming optimization problem:
\allowdisplaybreaks[0]
\begin{maxi*}
  {a_i \geq 0}{\sum_{k=0}^n a_k}{}{}
    \label{PLP}
  \addConstraint{\sum_{k=0}^n a_k K_s(k)}{\geq 0,\qquad}{s = 0, 1, \dots, n}\tag{$\mathcal{P}$}
  \vspace{-5pt}\addConstraint{a_0}{ = 1}{}
  \addConstraint{a_k}{ = 0}{1 \leq k \leq (d-1)}
\end{maxi*}
Denote $\bb(x) = \sum\limits_{k=0}^n \beta_k K_k(x)$. Then the the dual of the above LP can be written as
\begin{mini*}
  {\beta_i \geq 0}{\bb(0)}{}{}
  \label{LP}
  \addConstraint{\bb(u)}{\leq 0,\qquad}{u = d, d+1, \dots, n}\tag{$\mathcal{D}$}
  \addConstraint{\beta_0}{ = 1}{}
\end{mini*}
\allowdisplaybreaks
Denoting by $A_{LP}(n, d)$ the optimum values of the above two problems, the LP bound claims that $A(n, d) \leq A_{LP}(n, d)$. 

The optimal values of the LPs and asymptotic upper bound $R_{LP}(\delta ) = \limsup\limits_{n\to\infty} \frac1{n}\log A_{LP}(n, \lfloor\delta n\rfloor)$ are not known yet, however upper bounds on $A_{LP}(n, d)$ (and $R_{LP}(\delta )$) can be obtained by finding feasible solutions to the dual LP~\eqref{LP}. In particular, the (first) MRRW upper bound $R_{MRRW}(\delta) = H\left(\frac12 - \sqrt{\delta(1-\delta)}\right)$ is obtained in such a way in~\cite{MRRW} (in this paper we do not address the second MRRW bound).

To formulate the Delsarte's LP for the almost-balanced case, we point out that the variables $a_i$, $i = 0, 1, \dots, n$ in the  primal form of LP~\eqref{PLP}  correspond to the distance distribution of the code $C$, i.e. $a_i = \frac1{|C|}\left\lvert \{ (x, y) \in C\, |\, \Delta(x, y) = i\} \right\lvert$. The first set of constraints in the LP \eqref{PLP} then correspond to Delsarte-MacWilliam's inequalities for the transformed (dual) distance distribution, while the constraints $a_k = 0$ for $1 \leq k < d$ denote that there are no codewords at distance below $d$.

For an almost-balanced code $C$ there is no two codewords in $C$ at distance below $d$ or above $(n-d)$, i.e. $a_k = 0$ for $k < d$ or $k > (n-d)$. Therefore, the Delsarte's linear program for almost-balanced binary codes with distance $d$ is
\allowdisplaybreaks[0]
    \begin{maxi*}
  {a_i \geq 0}{\sum_{k=0}^n a_k}{}{}
  \label{bal_PLP}
  \addConstraint{\sum_{k=0}^n a_k K_s(k)}{\geq 0,\quad}{s = 0, 1, \dots, n}\tag{$\mathcal{BP}$}
  \addConstraint{a_0}{ = 1}{}
  \addConstraint{a_k}{ = 0}{1 \leq k \leq d-1 \ \text{and}\ k > (n-d).}
\end{maxi*}
\allowdisplaybreaks
The dual to the above is the following linear program:
\allowdisplaybreaks[0]
\begin{mini*}
  {\beta_i \geq 0}{\bb(0)}{}{}
  \label{bal_LP}
  \addConstraint{\bb(u)}{\leq 0,\quad}{u = d, d+1, \dots, n - d}\tag{$\mathcal{BD}$}
  \addConstraint{\beta_0}{ = 1}{\hspace{4.8cm}}
\end{mini*}
\allowdisplaybreaks
 (the difference from the LP~\eqref{LP} is that $\bb(u)$ for $u > (n-d)$ are no longer required to be non-positive.)

We denote by $B_{LP}(n, d)$ the optimal value of the above pair of LPs, and thus $B(n, d) \leq B_{LP}(n, d)$. In analogy with $R_{LP}$, define by
$\Rb_{LP}(\delta) = \limsup\limits_{n\to\infty}\frac1{n}\log B_{LP}(n, \lfloor\delta n\rfloor)$
the LP upper bound on the maximal asymptotic rate for almost-balanced codes of relative distance~$\delta$: $\Rb(\delta) \leq \Rb_{LP}(\delta)$.

\section{LP bound for almost-balanced codes with large distances}

In this section we find the exact optimal solutions to Delsarte's linear programs~\eqref{bal_LP} and~\eqref{bal_PLP} for almost-balanced codes with large distance.
Specifically, we prove

\begin{thm}
\label{thm:B_bounds} Let $n, d$ be even such that ${d \geq \frac{n - \sqrt{n}}{2} + 1}$. Then the optimal value of the linear programs~\ref{bal_LP} and~\ref{bal_PLP} is $B_{LP}(n, d) = \dfrac{4d(n-d)}{n - (n-2d)^2}$.
\end{thm}

The upper bound on the size of almost-balanced linear codes $B(n, d) \leq B_{LP}(n, d)$ for this regime is equivalent to the Grey-Rankin bound~\cite{grey, rankin, rankin2} on the size of self-complementary codes. This bound was also obtained by Delsarte by providing a \emph{feasible} solution to~\ref{bal_LP}, see, for instance, Problem (18) in Chapter 17 (p. 544) in~\cite{McWSloane}, or~\cite{Helleseth} for a more general case. What we show below is that this solution is in fact~\emph{optimal} for~\ref{bal_LP}, and therefore for~\ref{bal_PLP} via duality. 

This implies that the bound in Theorem~\ref{thm:B_bounds} is the best upper bound on $B(n, d)$ one can obtain using (a straightforward application of) Delsarte's LP, so the Grey-Rankin bound cannot be improved using this approach. It was already shown in the literature that the Grey-Rankin bound can sometimes be achieved~\cite{McGuire}, however the range of pairs $(n, d)$ for which this happens is limited, and is related to existence of designs with certain parameters.
Theorem~\ref{thm:B_bounds}, on the other hand, shows that the LP approach cannot improve the Grey-Rankin bound for any even $n, d$. Besides that, we think that finding optimal (instead of feasible) solutions to Delsarte's LP is of independent interest, even for this narrow regime and the special case of almost-balanced codes.

\smallskip
To prove Theorem~\ref{thm:B_bounds},
we start with the following claim
\begin{lem}
\label{lem:bal_bound}
Let $\theta = (\theta_0, \theta_1, \dots, \theta_n) \in \R^{n+1}$ be a feasible solution to the LP \eqref{bal_LP}. Define $\ot \in \R^{n+1}$ which coincides with $\theta$ on all even coordinates, and has $0$ elsewhere: $\ot = (\theta_0 , 0, \theta_2, 0 ,\theta_4, \dots)$. Then $\ot$ is a feasible solution to the LP~\eqref{bal_LP} for almost-balanced codes.
\end{lem}
\begin{IEEEproof}
Denote $\bt(x) = \sum\limits_{k=0}^n \theta_k K_k(x)$, then $\bt(u) \leq 0$ for any integer $u$ s.t. $d \leq u \leq (n-d)$. Consider then 
\[\obt(x) = \dfrac{\bt(x) + \bt(n-x)}{2} = \sum\limits_{k=0}^n \theta_k \dfrac{K_k(x) + K_k(n-x)}{2}.\]

By Fact~\ref{cor:parity}, $K_k(n/2 - y)$ is an even polynomial for even $k$, and so $\frac{K_k(x) + K_k(n-x)}{2} = K_k(x)$. On the other hand, $\frac{K_k(x) + K_k(n-x)}{2} = 0$ for odd $k$, and therefore ${\obt(x) =  \sum\limits_{k=0}^n \ot_k K_k(x)}$ by our definition of $\ot$.

For any integer $d \leq u \leq n-d$, we also have $d \leq (n-u) \leq (n-d)$. It means that $\bt(u) \leq 0$ and $\bt(n-u) \leq 0$, so $\obt(u) \leq 0$ for any such $u$. Therefore $\ot$ if feasible for \eqref{bal_LP}.
\end{IEEEproof}

\smallskip
\begin{cor}
\label{cor:opt_even_coords}
Optimal solution to~\eqref{bal_LP} has $\beta_i = 0$ for odd~$i$. 
\end{cor}
\smallskip
\begin{IEEEproof}
$\theta_i \geq 0$ and  $K_k(0) = \binom{n}{k} \geq 0$ in Lemma~\ref{lem:bal_bound}, thus $\obt(0) = \sum\limits_{k=0}^n \ot_k K_k(0) \leq \sum\limits_{k=0}^n \theta_k K_k(0) = \bt(0)$. So nullifying all the odd-indexed coordinates doesn't increase the objective value in~\ref{bal_LP}.
\end{IEEEproof}

\smallskip
Now, for every even distance $d$ such that $\frac{n - \sqrt{n}}{2} + 1 \leq d \leq n/2$ we find an optimal solution to~\eqref{bal_LP} and its dual~\eqref{bal_PLP}. The condition on $d$ comes from a restriction $|K_3(d)|\leq |K_2(d)|(n-2d)/3$, which we use in the proof and prove in Lemma~\ref{lem:K3d}.

Observe that $K_2(x) = 2x^2 - 2nx + \binom{n}{2}$ has roots $\frac{n\pm\sqrt{n}}{2}$, and so $K_2(d) < 0$ for distances $d$ of our interest. We obtain the optimal solution to~\eqref{bal_LP} with a {degree-$2$} function $\bb(x)$ (i.e. $\beta_i = 0$ for $i > 2$) such that $\beta_1 = 0$ (due to Corollary~\ref{cor:opt_even_coords}). Clearly then, the problem reduces to minimizing $\beta_2$ with the condition that $\bb(u) \leq 0$ for $u$ between $d$ and $(n - d)$. Since the function $\bb(x) = 1 + \beta_2K_2(x)$ is quadratic symmetric around $n/2$, it is clear that this is equivalent to the condition $\bb(d) = 1 + \beta_2K_2(d) \leq 0$. Thus the best degree-$2$ solution to~\eqref{bal_LP} is exactly $\beta_0 = 1$, $\beta_2 = -\frac1{K_2(d)}$, and $\beta_i = 0$ for all other~$i$. We now prove that this is actually the overall optimal solution to~\eqref{bal_LP} (for even $d$ and $n$).

\begin{lem}
\label{lem:optimal}
$\beta_0 = 1$, $\beta_2 = -\frac1{K_2(d)}$, and $\beta_i = 0$ for all other~$i$ is an optimal solution for~\eqref{bal_LP} when $\frac{n - \sqrt{n}}{2} + 1 \leq d \leq n/2$ and $n$, $d$ are even.
\end{lem}
\begin{IEEEproof}
We use duality of linear programming to prove optimally. Namely, we use the fact that if $\beta$ is feasible for the LP~$\eqref{bal_LP}$, some $\alpha$ is feasible for its dual LP~\eqref{bal_PLP}, and complementary slackness conditions are satisfied, then $\beta$ and $\alpha$ are optimal for the LP and its dual. 

Complementary slackness conditions for our case are:
\begin{equation}
    \begin{aligned}
    \bb(u)\cdot\alpha_u = 0 &\qquad&& u = d, d+1, \dots, (n-d),\\
    \beta_s\cdot \left(\sum_{k=0}^n \alpha_k K_s(k)\right)  = 0 &\qquad&& s = 1, 2, \dots, n.
    \end{aligned}
\end{equation}
Since $\bb(x) = 1 - K_2(x)/K_2(d)$ only has roots at $d$ and $(n-d)$, we immediately see that $\alpha_u = 0$ for all $u$ other than $d$ and $(n-d)$. Further, since $\beta_2 \not= 0$, we have that $\binom{n}{2} + \alpha_d\cdot K_2(d) + \alpha_{n-d}\cdot K_2(n-d) = 0$. We claim that taking \[\alpha_d = \alpha_{n-d} = -\frac{\binom{n}{2}}{2K_2(d)}\]
and $\alpha_i = 0$ for every other coordinate gives a feasible solution~$\alpha$ to the LP~\eqref{bal_PLP}. Notice that all the complementary slackness conditions are satisfied for such $\beta$ and $\alpha$.

We now prove that Delsarte-MacWilliam's inequalities in~\eqref{bal_PLP} are satisfied. Specifically, we need to show
\begin{equation}
    \label{eq:McW}
    \binom{n}{s} + \alpha_d K_s(d) + \alpha_{n-d} K_s(n-d) \geq 0, \qquad s = 0, 1, \dots, n.
\end{equation}

The case $s=0$ is straightforward as $\alpha \geq 0$, and for ${s = 2}$ the equality holds by the choice of $\alpha$. Next, for odd $s$, $K_s(d) = -K_s(n-d)$ by Fact~\ref{cor:parity}, and therefore 
$ \binom{n}{s} + \alpha_d\cdot K_s(d) + \alpha_{n-d}\cdot K_s(n- d) = \binom{n}{s} \geq 0$. Since $n$ is even, this applies to $s = (n-1)$. Moreover, for $s = n$, $K_n(d) = K_0(d)$ and $K_n(n-d) = K_0(n - d)$ (using Fact~\ref{cor:symmetry} and since $n, d$ are even), so the inequality also holds. 

For every even $s$ we have $K_s(d) = K_s(n-d)$. So it remains to prove for every even $2 < s < n - 1$ that
\begin{equation}
\label{eq:to_prove}    
\binom{n}{s} \geq -2\cdot\alpha_d\cdot K_s(d) = \frac{\binom{n}{2}}{K_2(d)} \cdot K_s(d).
\end{equation}
Denote for convenience $C = \left\lvert K_2(d)/\binom{n}{2} \right\lvert$. We prove the following statement, from which~\eqref{eq:to_prove} clearly follows
\begin{equation}
    \label{eq:stronger}
    |K_s(d)| \leq C\cdot \binom{n}{s}, \qquad s = 2, 3, \dots, (n-2).
\end{equation}
Notice that it is sufficient to show~\eqref{eq:stronger} only for $2 \leq s \leq n/2$, as the inequality for all other values of $s$ will follow from Fact~\ref{cor:parity}. 

We are going to need the following
\begin{lem}
\label{lem:recursive}
Let $|K_{q-1}(d)| \leq \delta\cdot \binom{n}{q-1}$ and $|K_{q}(d)| \leq \delta\cdot\binom{n}{q}$ for some $\delta > 0$, $d < n/2$, and positive integer $q < n$. Then
\begin{equation}
    |K_{q+1}(d)| \leq \delta\cdot\binom{n}{q+1}\cdot\frac{n-2d + q}{n-q}.
\end{equation}
\end{lem}

For clarity of exposition, we defer its proof until the end of this section.

We show~\eqref{eq:stronger} in two steps. First, we prove the following:

\noindent \textbf{Hypothesis:} For every $1 \leq t \leq d/2$,
\begin{equation}
\label{eq:induction}
\begin{aligned}
|K_{2t}(d)| &\leq C\cdot \binom{n}{2t} \cdot \prod_{i = 2}^t\frac{n - 2d + 2i}{n - 2i}, \\
|K_{2t + 1}(d)| &\leq C\cdot \binom{n}{2t + 1} \cdot \prod_{i = 2}^t\frac{n - 2d + 2i}{n - 2i},
\end{aligned}
\end{equation}
where empty products are treated as $1$.

\noindent \textbf{Base:} For $t=1$, we already know $|K_2(d)| = C\cdot \binom{n}{2}$. The proof of above inequality for $|K_3(d)|$ is defered to Lemma~\ref{lem:K3d}.

\noindent \textbf{Step:} Denote $\eta_t = \prod\limits_{i = 2}^t\frac{n - 2d + 2i}{n - 2i}$ for brevity. 
So, suppose~\eqref{eq:induction} holds for some $(t - 1)$, where $1 \leq t \leq d/2$. 
First, we derive from Lemma~\ref{lem:recursive} for $q = 2t - 1$:
\begin{equation}
    |K_{2t}(d)| \leq C\cdot\eta_{t-1}\cdot\binom{n}{2t}\cdot\frac{n-2d + (2t-1)}{n-(2t-1)} \leq C\cdot\eta_t\cdot\binom{n}{2t},
\end{equation}
where we use $\frac{n-2d+(2t-1)}{n-(2t-1)} \leq \frac{n-2d+2t}{n-2t}$. Notice also that $\frac{n-2d+2t}{n-2t} \leq 1$ for $t \leq d/2$, and so $\eta_t \leq \eta_{t-1}$. Then apply Lemma~\ref{lem:recursive} for $q = 2t$ and $\delta = C\cdot\eta_{t-1}$ again: 
\begin{equation}
    |K_{2t+1}(d)| \leq C\eta_{t-1}\cdot\binom{n}{2t+1}\frac{n-2d + 2t}{n-2t} = C\eta_t\binom{n}{2t + 1}.
\end{equation}

Therefore,~\eqref{eq:induction} holds for any $1 \leq t \leq d/2$. Denote $\Phi = \eta_{d/2} = \prod\limits_{i = 2}^{d/2}\frac{n - 2d + 2i}{n - 2i}$. We now prove the following

\noindent \textbf{Hypothesis:} For every $s$ such that $d \leq s \leq n/2$,
\begin{equation}
\label{eq:hypothesis}
    |K_s(d)| \leq C\cdot\Phi\cdot\binom{n}{s}\cdot\prod_{k=d+1}^{s-1}\frac{n - 2d + k}{n-k}.
\end{equation}

\noindent \textbf{Base:} The cases $s=d, (d+1)$ follow from~\eqref{eq:induction} for $t = d/2$.

\noindent \textbf{Step:} Denote $\mu_s = \prod\limits_{k=d+1}^{s-1}\frac{n - 2d + k}{n-k}$, and notice that $\mu_{s-1} \leq \mu_{s}$ for any $s$ within the range of interest. Suppose~\eqref{eq:hypothesis} holds for $(s-2)$ and $(s-1)$, and $(d + 2) \leq s \leq n/2$. Then apply Lemma~\ref{lem:recursive} for $q = (s-1)$ and $\delta = C\cdot\Phi\cdot\mu_{s-1}$\,:
\begin{equation}
    |K_s(d)| \leq C\Phi\cdot\mu_{s-1}\cdot\binom{n}{s}\frac{n-2d + (s-1)}{n-(s-1)} =  C\cdot\Phi\cdot\mu_{s}\cdot\binom{n}{s}.
\end{equation}

We are finally ready to prove~\eqref{eq:stronger} for every $s$ between $2$ and $n/2$. For $s$ such that $2\leq s \leq (d+1)$,~\eqref{eq:induction} implies $|K_s(d)| \leq C\cdot\eta_{\lfloor s/2\rfloor}\cdot\binom{n}{s}$. Clearly $\eta_{\lfloor s/2\rfloor} \leq 1$ as every term in the product is at most $1$, so~\eqref{eq:stronger} holds for such $s$.

Next, we have from~\eqref{eq:hypothesis} that $|K_s(d)| \leq C\cdot\Phi\cdot\mu_s\cdot\binom{n}{s} \leq C\cdot\Phi\cdot\mu_{n/2}\cdot\binom{n}{s}$ for every s between $(d + 2)$ and $n/2$, as $\mu_s$ is an increasing sequence. So it is sufficient to show that $\Phi\cdot\mu_{n/2} \leq 1$. Denote $w = n/2 - d - 1$, so $w < \sqrt{n}/2$, since $d > (n - \sqrt{n})/2$. Recall that $\Phi= \eta_{d/2} \leq \eta_{w+1}$, and so
\begin{align}
   \Phi\cdot\mu_{n/2} 
   & \leq \prod_{i=2}^{w + 1}\frac{n-2d+2i}{n-2i}\cdot\prod_{v=1}^{w}\frac{n-d+v}{n - d -v} \\
    & \leq \left(\frac{n-2d+2w+2}{n-2w-2}\right)^w\left(\frac{n-d+w}{n-d-w}\right)^w \\
   & = \left(\frac{4w+4}{n-2w-2}\cdot\frac{n/2 + 2w + 1}{n/2 + 1}\right)^w \leq 1.
\end{align}
 The final inequality clearly holds because $w < \sqrt{n}/2$ .
 
 This completes the proof of~\eqref{eq:stronger} for the whole range of $2 \leq s \leq (n-2)$, and together with our arguments that~\eqref{eq:to_prove} holds for $s \in \{0, 1, (n-1), n\}$, this means that all Delsarte-McWilliam's inequalities from~\eqref{bal_PLP} are satisfied. Therefore, we conclude that $\alpha$ is optimal for~\eqref{bal_PLP} and $\beta$ is optimal for~\eqref{bal_LP}.
\end{IEEEproof}

\begin{IEEEproof}[Proof of Lemma~\ref{lem:recursive}]
We use the following recurrence for Krawtchouk polynomials:
\begin{equation}
\label{eq:K_recur}
    (q + 1)K_{q+1}(x) = (n - 2x)K_q(x) - (n - q + 1)K_{q-1}(x),    
\end{equation}
for any positive integer $q < n$, where $K_0(x) = 1$, and $K_1(x) = n - 2x$. The proof now follows from a simple inductive calculation. \end{IEEEproof}

\begin{lem} Let $\frac{n-\sqrt{n}}{2} + 1 \leq d \leq n/2$. Then
\label{lem:K3d}
\begin{equation}
    \label{eq:K3d}
    |K_3(d)| \leq\frac{|K_2(d)|}{\binom{n}{2}}\binom{n}{3}= \frac{|K_2(d)|\cdot(n-2)}{3}.
\end{equation}
\end{lem}
\begin{IEEEproof}
We know $K_1(d) \geq 0$ and $K_2(d) < 0$ for such $d$. Using recurrence~\eqref{eq:K_recur} and this sign information, obtain
\begin{align}
    |K_3(d)| = \frac{-(n-2d)K_2(d) + (n-1)(n-2d)}{3}
\end{align}
We need to show that the above is at most $-\frac{K_2(d)\cdot(n-2)}{3}$. Equivalently, we need to prove 
\begin{equation}
    \label{eq:lem_K2d}
-2(d - 1)K_2(d) \overset{?}{\geq} (n-2d)(n-1).
\end{equation}
Decompose $K_2(d) = 2\left(d - \frac{n-\sqrt{n}}{2}\right)\left(d - \frac{n+\sqrt{n}}{2}\right)$, and using the conditions on $d$, we have $-K_2(d) \geq \sqrt{n}$.

Further, using $\sqrt{n} - 2 \geq (n - 2d)$
for~\eqref{eq:lem_K2d} we finally get
\begin{align}
-2(d - 1)K_2(d) \geq (n - \sqrt{n})\sqrt{n} &> (\sqrt{n} - 2)n\\
&\geq (n-2d)(n-1).\hspace{2.5em plus 1fill} \IEEEQEDhere
\end{align}
\end{IEEEproof}

This finally brings us the the proof of our main result.
\begin{IEEEproof}[Proof of Theorem~\ref{thm:B_bounds}] Observe that $K_2(d) = 2d^2 - 2nd - \binom{n}{2} = -\frac12\left(n - (n-2d)^2\right)$. Applying Lemma~\ref{lem:optimal} obtain
\begin{align}
B_{LP}(n, d) = 1 - \frac{\binom{n}{2}}{K_2(d)} &= 1 + \frac{n(n-1)}{n - (n-2d)^2} \\
& = \frac{4d(n-d)}{n - (n-2d)^2}. \hspace{6.5em plus 2fill}\IEEEQEDhere
\end{align}
\end{IEEEproof}

\section{Lower bound on LP bound for almost-balanced codes}
 Consider asymptotic lower and upper bounds on $R(\delta)$ \[R_{GV}(\delta) \leq R(\delta) \leq R_{MRRW}(
 \delta),\]
where $R_{GV}(\delta) = 1 - H(\delta)$ is a Gilbert-Varshamov bound obtained using a standard packing argument, which is currently the best known lower bound on $R(\delta)$. The bound $R_{MRRW}(\delta) = H\left(\frac12 - \sqrt{\delta(1-\delta)}\right)$ is the first MRRW~\cite{MRRW} bound, which is the best known upper bound for $\delta > 0.273$ (the second MRRW bound  is the best known for the remaining range of $\delta$).

In~\cite{Samorodnistky} Samorodnitsky proved an integrality gap of at most~$2$ for the MRRW bound with respect to the true LP bound:
    \begin{equation}
    \label{eq:Samor}
    \begin{aligned}  
    \dfrac{R_{GV}(\delta) + R_{MRRW}(\delta)}{2}\leq R_{LP}(\delta) \leq R_{MRRW}(\delta).
    \end{aligned}
    \end{equation}
 Combined    with the fact that $R_{MRRW}(\delta) > R_{GV}(\delta)$ for any $\delta \in (0, \frac12)$, the above proved that Delsarte's linear programming bound cannot attain the currently best known lower bound on $R(\delta)$.

In this section we prove an analogous result for the linear programming bound for the almost-balanced codes. Our proof is a slight modification of a proof from \cite{Samorodnistky}.

\vspace{4pt}
\begin{thm}
\label{thm:bal_Samor}
 $\Rb_{LP}(\delta) \geq \dfrac{R_{GV}(\delta)\hspace{-3pt} +\hspace{-3pt} R_{MRRW}(\delta)}{2}$ for any ${\delta\hspace{-1pt} \in \hspace{-1pt}(0, \frac12)}$.
\end{thm}

 To obtain a lower bound, we derive a feasible solution to \eqref{bal_PLP}, closely following \cite{Samorodnistky}.

\begin{lem}
\label{lem:bal_Samor}
 Let $\e = \dfrac1{4n}\sqrt{\dfrac{\binom{n}{\lfloor x_d\rfloor}}{2^n\cdot\binom{n}{d}}}$, where $x_d$ is the first (smallest) root of the polynomial $K_d(x)$, and let
 \begin{itemize}
     \item $a_0 = 1$
     \item $a_k = 0$ for $0 \leq k \leq d-1$ and ${(n - d + 1)} \leq k \leq n$
     \item $a_d = a_{n-d} = \e\cdot (d+1)\cdot \binom{n}{d}$
     \item $a_k = \e\cdot\binom{n}{k}$ for $d < k < (n - d)$.
 \end{itemize} 
 Then $a_0, \dots, a_n$ is a feasible solution to the LP~\eqref{bal_PLP}. 
\end{lem}
\begin{IEEEproof}
Clearly, we only need to verify the first set of constraints in \eqref{bal_PLP} (Delsarte-MacWilliam's inequalities). The case $s = 0$ is immediate, so assume $s \geq 1$. 
We have
\begin{align}
\label{eq:pf_n1}
    \sum_{k=0}^n a_k K_s(k) = &K_s(0) + \e\cdot\sum_{k = d}^{n - d}\binom{n}{k}K_s(k) \\ 
    \label{eq:pf_n2}
    &\hspace{20pt}+ \e\cdot d\binom{n}{d}\Big(K_s(d) + K_s(n - d)\Big) \\ 
    & \hspace{-65pt}= \binom{n}{s} + \e\sum_{k = d}^{n - d}\binom{n}{s}K_k(s) + \e d\binom{n}{s}\Big(K_d(s) + K_{n-d}(s)\Big),
\label{eq:pf0}
\end{align}
where we used reciprocity property from Fact~\ref{cor:reciprocity}. 


Next we use Facts~\ref{cor:parity}-\ref{cor:symmetry}, and consider two cases. When $s$ is odd, everything except the first summand in the RHS of \eqref{eq:pf0} cancels out. Indeed, $K_d(s) = - K_{n-d}(s)$, and the summation in the middle can be written as 
\begin{equation}
\begin{aligned}\sum_{k = d}^{n - d}\binom{n}{s}K_k(s) = &\binom{n}{s}\sum_{k = d}^{\frac{n}{2} - 1}\big(K_k(s) + K_{n - k}(s)\big)\\
& + \binom{n}{s}\cdot K_{n/2}(s)\cdot \big[(n+1)\ \text{mod}\  2\big],
\end{aligned}
\end{equation}
Observe that $K_k(s) + K_{n - k}(s) = 0$ for any $k$ within the summation range. Finally, for even $n$ we have $\binom{n}{s} K_{n/2}(s) = \binom{n}{n/2}K_s(n/2) = 0$, as $K_s(n/2 - x)$ is an odd function. Therefore, 
$\sum\limits_{k=0}^n a_k K_s(k) = K_s(0) = \binom{n}{s} \geq 0$ for odd $s$. 

Now consider the case of even $s$. Using the fact that Krawtchouk polynomials are orthogonal with respect to the binomial measure $\mu(k) = \binom{n}{k}/2^n$ and that $K_0(k) \equiv 1$, obtain $2^n\cdot\sum_{k=0}^n \mu(k)K_s(k)\cdot K_0(k) = \sum_{k=0}^n \binom{n}{k}K_s(k) = 0$. Then in~\eqref{eq:pf_n1}-\eqref{eq:pf_n2} for the summation in the RHS we can write 
\begin{equation}
\begin{aligned}
\sum_{k = d}^{n - d}&\binom{n}{k}K_s(k) = - \sum_{k = 0}^{d-1}\binom{n}{k}K_s(k) - \hspace{-12pt} \sum_{k = n - d + 1}^{n}\hspace{-4pt}\binom{n}{k}K_s(k) \\
& = - \sum_{k = 0}^{d-1}\binom{n}{s}K_k(s) - \hspace{-12pt} \sum_{k = n - d + 1}^{n}\hspace{-4pt}\binom{n}{s}K_k(s)   \\
& = - \binom{n}{s} \hspace{-2pt}\sum_{k=0}^{d-1}\hspace{-2pt}\Big(K_k(s) + K_{n-k}(s)\Big) =  - 2\binom{n}{s}\hspace{-2pt} \sum_{k=0}^{d-1} K_k(s),
\end{aligned}
\end{equation}
where we used reciprocity and Fact~\ref{cor:symmetry} for even $s$. Using these properties again for last part of RHS in~\eqref{eq:pf_n1}-\eqref{eq:pf_n2}, we obtain
\begin{equation}
    \sum_{k=0}^n a_k K_s(k) = \binom{n}{s}\left(1 - 2\e\sum_{k=0}^{d-1}K_k(s) + 2\e\cdot d\, K_d(s) \right).
\end{equation}

Finally, we notice that the RHS of the above equation is exactly the expression derived by Samorodnitsky in \cite[eq.~(22)]{Samorodnistky} (we took $\e$ exactly two times smaller than in \cite{Samorodnistky} for these expressions to coincide), where it was proven to be non-negative. Therefore, $a_0, \dots, a_n$ is feasible for \eqref{bal_PLP}.
\end{IEEEproof}


\begin{IEEEproof}[Proof of Theorem~\ref{thm:bal_Samor}] Taking the feasible solution $a_0, \dots, a_n$ for \eqref{bal_PLP} from Lemma~\ref{lem:bal_Samor} we obtain
\[ B_{LP}(n, d) \geq \sum_{k=0}^na_k \geq \e\sum_{k=d}^{n-d}\binom{n}{k}.\]
Consider some fixed $\delta \in (0, 1/2)$ and $d = \lfloor\delta n\rfloor$ as $n$ increases. Standard concentration properties of Binomial distribution (e.g. Chernoff bound) then imply that for large enough $n$, most of the weight will lie between $\binom{n}{d}/2^n$ and $\binom{n}{n-d}/2^n$. Then for such large $n$ we write
\[ B_{LP}(n, d) \geq \e\cdot 2^{n-1} \geq \dfrac1{8n}\sqrt{\dfrac{\binom{n}{\lfloor x_d\rfloor}\cdot 2^n}{\binom{n}{d}}}.\]

Finally, we use an asymptotic for the first root of Krawtchouk polynomial $K_d$ as $n$ goes to infinity: ${x_d = n\left(\frac12 - \sqrt{\frac{d}{n}\left(1 - \frac{d}{n}\right)}\right) + o(n)}$. Together with an asymptotic $\lim\limits_{n\to\infty}\frac1{n}\log_2\binom{n}{\gamma n} = H(\gamma)$, we derive
{\small
\begin{equation*}
\begin{aligned}
&\Rb_{LP}(\delta) \geq \lim\limits_{n\to\infty}\frac1{2n}\Bigg[ \log_2\binom{n}{\lfloor x_d\rfloor} + n - \log_2\binom{n}{\delta n} \Bigg] \\
\\[-10pt] 
&= \dfrac{1 - H(\delta) + H\left(1/2 - \sqrt{\delta\left(1 - \delta\right)}\right)}{2} = \dfrac{R_{GV}(\delta) + R_{MRRW}(\delta)}{2}.
\end{aligned}
\end{equation*}
}
\end{IEEEproof}

\section*{Acknowledgment}
Research supported in part by NSF grants CCF-1563742 and CCF-1814603.

We thank the anonymous reviewers who provided valuable comments about the paper and brought important references to our attention.

\newpage
\IEEEtriggeratref{5}
\bibliographystyle{IEEEtran}
\bibliography{bib}

\end{document}